\documentclass[aip,jcp,preprint,superscriptaddress]{revtex4-1}

\usepackage{amsmath}
\usepackage{graphicx,color}
\usepackage[update,prepend]{epstopdf}
\usepackage{enumitem}

\begin{document}
\title{Influence of hydrodynamic interactions on stratification in drying mixtures}

\author{Antonia Statt}
\thanks{A. Statt and M.P. Howard contributed equally to this work.}
\affiliation{Department of Chemical and Biological Engineering, Princeton University, Princeton, New Jersey 08544, USA}

\author{Michael P. Howard}
\thanks{A. Statt and M.P. Howard contributed equally to this work.}
\affiliation{Department of Chemical and Biological Engineering, Princeton University, Princeton, New Jersey 08544, USA}

\author{Athanassios Z. Panagiotopoulos}
\email{azp@princeton.edu}
\affiliation{Department of Chemical and Biological Engineering, Princeton University, Princeton, New Jersey 08544, USA}

\newcommand{\vv}[1]{\mathbf{#1}}

\begin{abstract}
Nonequilibrium molecular dynamics simulations are used to investigate the influence of hydrodynamic interactions on vertical
segregation (stratification) in drying mixtures of long and short polymer chains. In agreement with previous computer
simulations and theoretical modeling, the short polymers stratify on top of the long polymers at the top of the drying film
when hydrodynamic interactions between polymers are neglected. However, no stratification occurs at the
same drying conditions when hydrodynamic interactions are incorporated through an explicit solvent
model. Our analysis demonstrates that models lacking hydrodynamic interactions do not faithfully
represent stratification in drying mixtures, in agreement with recent analysis of an idealized model for diffusiophoresis,
and must be incorporated into such models in future.
\end{abstract}
\maketitle

\section{Introduction}
Coatings or films formed by drying\cite{Russel:2011fl,Routh:2013dm} are relevant to many technologies,
including latex paints,\cite{Keddie:1997vt,Keddie:2010up} inkjet printing,\cite{Calvert:2001ez} and
polymer nanocomposites.\cite{Kumar:2017wz} Such films are often comprised of multiple
components including nanoparticles or colloids,\cite{Russel:1989} polymers,\cite{Padget:1994wq} and
surfactants.\cite{Keddie:2010up,Hellgren:1999uv} It is desirable to control the distribution of components within the
dried film, e.g., to ensure the dispersion of nanoparticles\cite{Kumar:2017wz,Mackay:2006bg} or to
impart antimicrobial properties to a surface.\cite{Fulmer:2011iq}
Recent experiments showed that quickly dried mixtures of small and large spherical colloids can vertically
stratify into layers.\cite{Fortini:2016ip,MartinFabiani:2016fj,Makepeace:2017ht} Unexpectedly, the smaller colloids
accumulated at the solvent-air interface and pushed the larger colloids down into the film. Computer
simulations qualitatively agreed with the experiments,\cite{Fortini:2016ip,MartinFabiani:2016fj,Makepeace:2017ht,Howard:2017bq}
and formed the same small-on-top stratification in ternary and polydisperse colloid mixtures\cite{Fortini:2017jl}
and in polymer-polymer and colloid-polymer mixtures.\cite{Howard:2017vu} The stratification
was more pronounced for larger size ratios and faster drying speeds, although it was found to persist even at
moderate drying rates.\cite{Howard:2017bq,MartinFabiani:2016fj,Howard:2017vu}

Stratification has been modeled theoretically as a diffusion process\cite{Russel:1989,Routh:2004jz} characterized by
the film P\'{e}clet number for each component, ${\rm Pe}_i = H v/D_i$, where $H$ is the initial film height,
$v$ is the speed of the solvent-air interface, and $D_i$ is the diffusion coefficient of component $i$.
When ${\rm Pe}_i \ll 1$, diffusion dominates over advection from drying, and a nearly uniform distribution of component $i$
is expected. When ${\rm Pe}_i \gg 1$, component $i$ accumulates toward the drying interface. Trueman et al. developed a model
for drying colloid mixtures\cite{Trueman:2012er} that qualitatively predicted a maximum in the number of
large colloids on top of the film when ${\rm Pe}_1 < 1$ and ${\rm Pe}_2 > 1$ (here 1 is the smaller colloid and 2
is the larger colloid).\cite{Trueman:2012ve,Liu:2018vb} However, their model does not predict the small-on-top
stratification that occurs when ${\rm Pe}_1 > 1$ and ${\rm Pe}_2 > 1$.\cite{Trueman:2012er,Trueman:2012ve} Fortini et al.
hypothesized that such stratification resulted from an osmotic pressure imbalance,\cite{Fortini:2016ip,Fortini:2017jl}
while Howard et al.\cite{Howard:2017bq,Howard:2017vu} and Zhou et al.\cite{Zhou:2017gg} proposed diffusion models using
the chemical potential as the driving force. Computer simulations of polymer mixtures and the Howard et al. model
agreed quantitatively,\cite{Howard:2017vu} and qualitative agreement was found between the Zhou et al. model and
experiments for colloid mixtures in certain parameter regimes.\cite{Makepeace:2017ht,Liu:2018vb} The Zhou et al. model
is equivalent to the low-density limit of the Howard et al. model.\cite{Howard:2017bq}

For computational and theoretical convenience, many previous studies have employed a free-draining approximation for hydrodynamic
interactions on the colloids or polymers, incorporating the Stokes drag on each particle but neglecting other interactions.
The computer simulations\cite{Fortini:2016ip,MartinFabiani:2016fj,Makepeace:2017ht,Howard:2017bq,Fortini:2017jl,Howard:2017vu}
implicitly modeled the solvent using Langevin or Brownian dynamics methods that treat the solvent as a quiescent,
viscous background, while the theoretical models\cite{Fortini:2016ip,Fortini:2017jl,Zhou:2017gg,Howard:2017bq,Howard:2017vu}
assumed that the particle mobilities were given by the Einstein relation. However, Sear and Warren
point out in a recent analysis\cite{Sear:2017cv} that neglecting hydrodynamic interactions overpredicts the phoretic velocity of
a single large colloid in an ideal polymer solution, which must be proportional to the extent of stratification in this simple model.
They argue that this discrepancy is caused by the omission of solvent backflow,\cite{Sear:2017cv}
leading to an incorrect hydrodynamic mobility.\cite{Brady:2011bh} Their argument is qualitatively supported by differences
between the experiments and implicit-solvent simulations of Fortini et al.\cite{Fortini:2016ip}
To our knowledge, though, there has been no direct test of how hydrodynamic interactions influence stratification
at finite concentrations or size ratios between components, which are more challenging to analyze theoretically.\cite{Sear:2017cv,Brady:2011bh}

In this article, we systematically demonstrate the influence of hydrodynamic interactions on stratification for a drying polymer mixture.
Computer simulations are ideal tools for such a study because, unlike in experiments, it is possible to artificially remove
hydrodynamic interactions between particles in the simulations. Molecular dynamics simulations of drying were performed using both
explicit and implicit solvent models. We show that although the implicit-solvent model stratifies for the drying conditions
investigated, the explicit-solvent model does not due to the presence of hydrodynamic interactions. Our simulations show
that it is critical to incorporate hydrodynamic interactions into models and simulations in order to reliably predict
stratification in drying mixtures.

\section{Models and Methods \label{sec:model}}
We performed nonequilibrium molecular dynamics simulations of drying polymer mixtures using two bead-spring models: one with an
explicitly resolved solvent (\ref{sec:explicitmodel}) and a corresponding one with an implicit solvent (\ref{sec:implicitmodel}).
While the explicit model included full hydrodynamic interactions, the implicit model
only incorporated the Stokes drag on each polymer, thus omitting hydrodynamic interactions between polymers.
Details of both models are presented next.

\subsection{Explicit solvent model \label{sec:explicitmodel}}
Polymers were represented as linear chains of monomers with diameter $\sigma$ and mass $m$ immersed in an explicit solvent
of beads of equal size and mass. All monomers and solvent particles interacted with the Lennard-Jones potential
\begin{equation}
U_\text{nb}(r) = 4\varepsilon \left[ \left(\frac{\sigma}{r}\right)^{12} - \left(\frac{\sigma}{r}\right)^{6} \right]\label{eq:wca}
\end{equation}
where $\varepsilon$ is the interaction energy and $r$ is the distance between particles. The monomer--monomer interactions
were made purely repulsive by truncating the potential at its minimum and shifting it to zero.\cite{WEEKS:1971uq}
The solvent--solvent and monomer--solvent interactions included the attractive part of $U_{\rm nb}$ and were truncated
at  $3.0\,\sigma$  with the energy and forces smoothed to zero starting from $2.5\,\sigma$.
Bonds between monomers were represented by the finitely extensible nonlinear elastic potential~\cite{GREST:1986wy}
\begin{equation}
U_\text{b}(r) = \begin{cases}
-\dfrac{1}{2}\kappa r_0^2 \ln\left[1 - \left(\dfrac{r}{r_0}\right)^2 \right], &  r < r_0\\
\infty ,&  r \geq r_0 \\
\end{cases}
\label{eq:FENE}
\end{equation}
with the standard parameters $r_0 =1.5\sigma$ and $\kappa = 30\,\varepsilon/\sigma^2$. This choice prevents unphysical
bond crossing because the equilibrium bond length is approximately $0.97\,\sigma$ at a temperature of
$T = 1.0\,\varepsilon/k_\text{B}$, where $k_\text{B}$ is Boltzman's constant. This model corresponds to
good solvent conditions for the polymers.

The drying process was simulated using the method developed by Cheng and coworkers.\cite{Cheng:2011,Cheng:2013,Cheng:2016}
The polymer solution was supported by a smooth, structureless substrate modeled by a Lennard-Jones 9-3 potential at the
bottom of the simulation box
\begin{equation}
U_{\rm w}(z) = \varepsilon_{\rm w} \left[ \frac{2}{15}\left(\dfrac{\sigma}{z}\right)^9 - \left(\frac{\sigma}{z}\right)^3 \right] \label{eq:wall}
\end{equation}
where $z$ is the distance between the particle and the substrate, and $\varepsilon_{\rm w} = 2\,\varepsilon$ is the strength of the interaction.
Interactions were truncated for $z > 3.0\,\sigma$. The solvent vapor above the liquid film was confined by an additional potential of the same
form as $U_{\rm w}$ at the top of the simulation box, which was made purely repulsive by truncating it for $z > (2/5)^{1/6}\,\sigma$.
Solvent was evaporated by deleting a small number of randomly-chosen solvent particles from the top $20\,\sigma$ of the vapor.
To maintain temperature control, monomers and solvent in a slab of height $20\,\sigma$
above the substrate were weakly coupled to a Langevin thermostat with $T= 1.0\,\varepsilon/k_\text{B}$
and friction coefficient $0.1\,m/\tau$,\cite{Schneider:1978ge,Allen:1991,Phillips:2011td}
where $\tau = \sqrt{m\sigma^2/\varepsilon}$ is the derived unit of time.

All simulations were performed using the HOOMD-blue simulation package (version 2.2.2) on multiple graphics processing
units~\cite{Anderson:2008vg,Glaser:2015cu} with a timestep of $\Delta t = 0.005\,\tau$. The solvent coexistence densities
were $0.664 \,\sigma^{-3}$ and $0.044\,\sigma^{-3}$ for the liquid and vapor phases, respectively. The simulation box was
periodic in the $x$ and $y$ directions with a length of $L = 50\,\sigma$. The height of the box was $L_z = 540\,\sigma$,
where $H=500\,\sigma$ was the initial film height and the remaining $40\,\sigma$ was filled with solvent vapor.
We investigated a mixture of short polymers of $M_1=10$ monomers and long polymers of $M_2=80$ monomers at an initial
monomer density of $0.006 \,\sigma^{-3}$ each.
The simulations consisted of 730000 solvent particles, 900 long chains, and 7100 short chains, resulting in a total
of $873000$ particles. By deleting one solvent particle every $1.25\tau$, we obtained an evaporative flux
of $3.2 \times 10^{-4} /\sigma^2\tau$.

\subsection{Implicit solvent model\label{sec:implicitmodel}}
The implicit-solvent model was constructed by matching the structure of the polymer chains in the explicit solvent at
infinite dilution. Interactions between the monomers and the solvent swelled the polymers beyond the size of a
chain with only the monomer--monomer interactions, so additional effective interactions were required.
We measured the distance distribution between the first and third monomer along a chain surrounded by solvent and a
chain without solvent. We fit the negative logarithm of the ratio of the explicit and implicit distribution with a
spline potential~\cite{Howard:2017} of the form
\begin{equation}
U_{\rm s}(r) = \begin{cases}
\varepsilon_{\rm s}, & r < r_{\rm s} \\
\varepsilon_{\rm s}(r_{\rm c}^2 - r)^n\dfrac{r^2_{\rm c} + n r^2 - (n+1)r_{\rm s}^2}{(r_{\rm c}^2 - r_{\rm s}^2)^{n+1}}, & r_{\rm s} \le r < r_{\rm c} \\
0, & r \ge r_{\rm c}
\end{cases}
\end{equation}
The fit resulted in parameters $n=3.8$, $ \varepsilon_s=0.68\,\varepsilon$, $r_{\rm s}=1.32\,\sigma$, and $r_{\rm c}=2.0\,\sigma$.
This potential, modeling the soft repulsion between monomers due to the solvent, was added to the bare monomer--monomer
interactions. We validated this effective model by measuring the polymer structure over a range of concentrations and compositions
up to a total monomer number density of $0.375\,\sigma^{-3}$, finding overall good agreement between the implicit and
explicit models (see Supplementary Material).

The polymer long-time dynamics in the explicit solvent were matched in the implicit solvent using Langevin dynamics simulations.\cite{Schneider:1978ge,Allen:1991,Phillips:2011td}
This technique incorporates the effects of Brownian motion and Stokes drag from the solvent, but neglects hydrodynamic
interactions between monomers. The monomer friction coefficients were adjusted for each polymer to give
the same polymer center-of-mass diffusion coefficient at infinite dilution as in the explicit solvent.
This approach gives the correct long-time dynamics, but distorts the internal relaxation modes of the polymers.\cite{Howard:2017vu}
We measured diffusion constants of $D_1 = 0.0342\,\sigma^2/\tau$ for the $M_1=10$ polymers and
$D_2 = 0.0071\,\sigma^2/\tau$ for the $M_2=80$ polymers from the polymer center-of-mass mean squared displacement in a cubic box
with edge length $L=50\,\sigma$, giving $2.92\,m/\tau$ and $1.76\,m/\tau$ for the friction coefficients, respectively.

The liquid-vapor interface was modeled by the repulsive part of a harmonic potential.\cite{Howard:2017bq,Howard:2017vu}
(The complete form of the potential can be found in ref.~\citenum{Howard:2017bq}.) In order to closely match the
explicit-solvent simulations, the spring constant and position of the interface were adjusted to obtain similar initial
density profiles to the explicit-solvent model. We used spring constants of $0.1\,\varepsilon/\sigma^2$ and
$0.18\,\varepsilon/\sigma^2$ for the monomers of the short and long chains, respectively. The position of the
interface was measured throughout the explicit-solvent simulations and used directly in the implicit-solvent simulations.
The minimum of the potential was offset for the long polymers by $-1\,\sigma$.

\section{Results and discussion \label{sec:results}}
\begin{figure}[h]
	\centering
	\includegraphics[width=7.5cm]{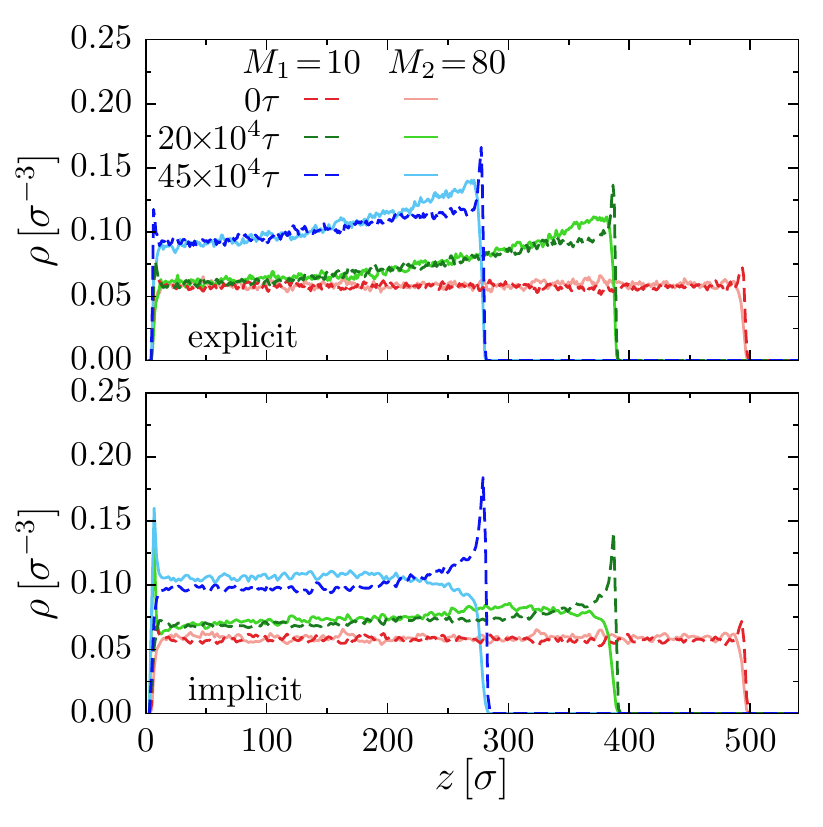}
	\caption{Monomer density profiles for the explicit-solvent model (top) and implicit-solvent model (bottom) for
	         three different times during evaporation. Solid lines show the profiles of the long ($M_2=80$) polymer,
	         while dashed lines are the profiles of the short ($M_1=10$) polymer.}
	\label{fig:density_profiles}
\end{figure}

We performed 25 drying simulations for both models and measured the monomer density profiles in the film.
Figure~\ref{fig:density_profiles} shows the average profiles at three times: before evaporation,
at a film height of roughly $3H_0/4$, and at a film height close to $H_0/2$. We used the average interface speed and the
diffusion coefficients at infinite dilution to estimate film P\'{e}clet numbers of $\text{Pe}_1 \approx 7$ and $\text{Pe}_2 \approx 33$.
(Throughout this discussion, 1 denotes the shorter polymer, while 2 is the longer polymer.)
Initially, the polymers were nearly uniformly distributed in the film.
When drying began, both the long and short polymers accumulated at the moving interface, as expected from their
P\'{e}clet numbers. However, there was a noticeable difference in the distribution of chains within the film between
the two models. Consistent with previous simulations and the Howard et al. theoretical model,\cite{Howard:2017vu}
the implicit-solvent model stratified with the shorter polymers on top of the longer polymers (Figure~\ref{fig:density_profiles}b).
In stark contrast, the explicit-solvent model showed no small-on-top stratification (Figure~\ref{fig:density_profiles}a),
and in fact more long polymers than short polymers accumulated immediately below the liquid-vapor interface.

The qualitatively different behavior of the two models suggests a significant difference in the relative migration speeds
of the polymer chains, which was previously shown to give rise to stratification for the implicit-solvent model.\cite{Fortini:2016ip,Howard:2017bq,Howard:2017vu,Zhou:2017gg}
At isothermal conditions, the relative migration speed $\Delta u$ predicted by the Howard et al. model\cite{Howard:2017bq} is
\begin{align}
\Delta u/u_1 = \frac{\mathcal{M}_2}{\mathcal{M}_1}\frac{(\partial \mu_2 / \partial z)}{(\partial \mu_1 / \partial z)} -1 \label{eq:migrate}
\end{align}
where $u_1$ is the migration speed of the short polymers, $\mu_i$ is the chemical potential of component $i$ computed
from the equilibrium free-energy functional, and $\mathcal{M}_i$ is the effective mobility for component $i$.
We previously estimated $M_i$ from the equilibrium diffusion coefficient $D_i$ using the Einstein relation,
$\mathcal{M}_i = D_i / k_{\rm B} T$, within the free-draining approximation of hydrodynamics.\cite{Howard:2017bq,Howard:2017vu}
Our simulations and equation~\ref{eq:migrate} then suggest several possible explanations
for the different drying behavior:
\begin{enumerate}[label=\Alph*.]
\item{temperature gradients from evaporative cooling affect $\mu_i$ or $\mathcal{M}_i$ in the explicit model, but are absent from the implicit model,}
\item{differences in the chemical potentials of the two models lead to different diffusive driving forces,}
\item{differences in the equilibrium diffusion coefficients change $\mathcal{M}_i$ for the two models, and/or}
\item{hydrodynamic interactions mediated by the explicit solvent alter $\mathcal{M}_i$ and the migration velocities.}
\end{enumerate}
We probed each of these effects in turn. As will be shown, we found that only hydrodynamic interactions (D) had a
significant enough effect to explain the lack of stratification in the explicit-solvent model.

\subsection{Temperature}
\begin{figure}[h]
	\centering
	\includegraphics[width=7.5cm]{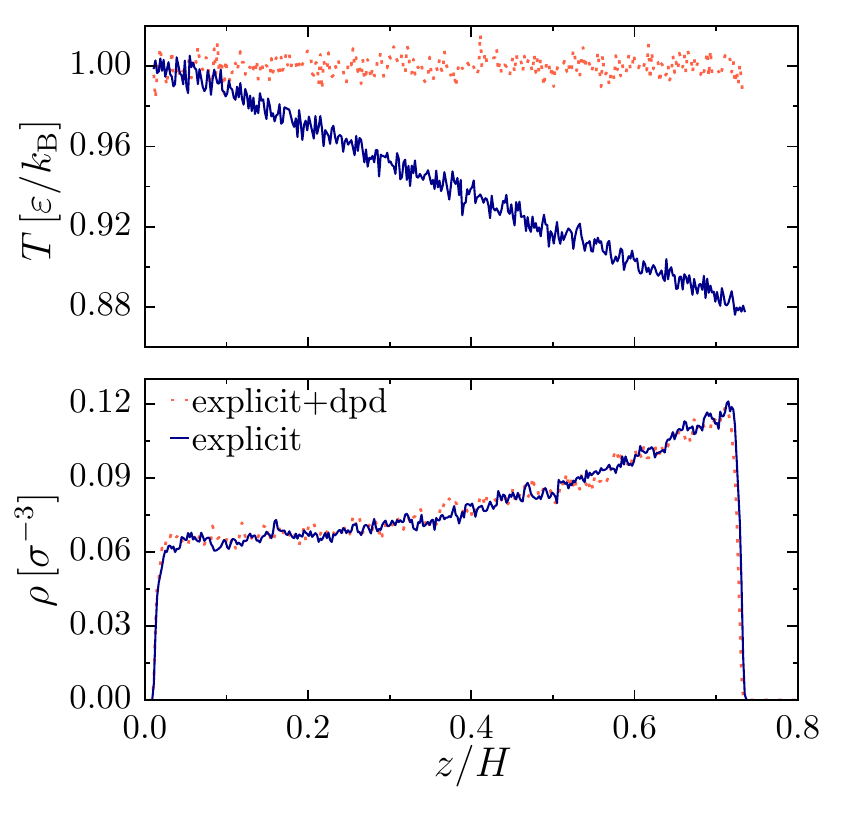}
	\caption{Temperature (top) and long polymer ($M_2 = 80$) monomer density (bottom) profiles in the explicit solvent
	         with (dashed line) and without (solid line) a DPD thermostat. The $z$ position is normalized by
	         the initial film height $H = 500\,\sigma$. Profiles are shown at $275000\, \tau$ with the thermostat
	         and $250000\, \tau$ without the thermostat.}
	\label{fig:temperature_gradient}
\end{figure}

Solvent evaporation can lead to cooling at fast drying rates.\cite{Cheng:2011} For our explicit-solvent model, the temperature
is expected to be fixed at $T = 1.0\,\varepsilon/k_{\rm B}$ close to the substrate and, at pseudo-steady state, to
decrease linearly in distance from the substrate with a slope proportional to the evaporative flux.
The temperature profile $T(z)$ measured in the simulations, shown in Figure~\ref{fig:temperature_gradient}a, is consistent
with this expectation. Here, the local temperature $T(z)$ is defined by
\begin{equation}
T(z) = \frac{m}{3 N_k k_\text{B}} \sum_{j = 1}^{N_k} v_j^2
\end{equation}
where the sum is taken over the $N_k$ particles in a slab $k$ centered around $z$ with thickness
$2\Delta z = 1\,\sigma$, and $\mathbf{v}_j$ is the velocity of particle $j$. On the other hand, there were no
temperature gradients in the implicit-solvent model because the solvent was treated as an isothermal, viscous background.

Temperature gradients can give rise to mass flux (i.e., the Soret effect),\cite{Reith:2000wp} and the mobility is
temperature dependent through the viscosity.\cite{Rowley:1997bi}
In order to test how the presence of temperature gradients influenced diffusion in the explicit-solvent model,
we coupled all particles to a dissipative particle dynamics (DPD) thermostat.\cite{Groot:1997du,Soddemann:2003cp,Phillips:2011td}
The DPD thermostat applies pairwise random and dissipative forces to all particles, and so conserves momentum and
preserves hydrodynamic interactions. We set the DPD friction coefficient to be $0.5\,m/\tau$, which has been
shown to have a negligible effect on the viscosity of a fluid of nearly-hard spheres.\cite{Soddemann:2003cp}
Figure~\ref{fig:temperature_gradient}a shows that although there was originally roughly 12\% evaporative cooling in the
explicit-solvent model, the DPD thermostat maintained a constant temperature throughout the film.

We then compared the distribution of polymers in the film with and without evaporative cooling.
The temperature gradient in the film led to a corresponding gradient in the the local total density (solvent and polymer),
which was higher in colder regions. This in turn decreased the total film height compared to the isothermal case
for equal amount of solvent evaporated. To account for this difference,
we compared the monomer density profiles at equal film height $H$. The profile for the $M_2=80$ polymers is shown in
Figure~\ref{fig:temperature_gradient}b for $H = 360\,\sigma$, when the temperature gradient was most pronounced.
The monomer density profiles were indistinguishable with and without evaporative cooling. Accordingly, we excluded
temperature effects as a possible reason for the different stratification behavior in the implicit- and explicit-solvent
models.

\subsection{Chemical potential \label{sec:chempot}}
In the absence of temperature gradients, diffusive flux is driven by chemical potential gradients within
the regime of linear response.\cite{deGroot:1984} A mismatch in the density or composition dependence of the chemical
potentials could result in different stratification behavior for the explicit- and implicit-solvent models.
We accordingly measured the relevant chemical potentials for both models using test insertion methods.
We first generated trajectories of polymer mixtures in cubic, periodic simulation boxes
over a wide range of mixture compositions (see Supplementary Material), saving 100 independent configurations
for each composition.

For the explicit-solvent mixtures, the chemical potential of the solvent $\mu_0$ was calculated by Widom's
insertion method\cite{Widom:1963gu}
\begin{equation}
\mu_0 = \mu_0^{\rm id} - k_{\rm B} T \ln \langle e^{-\beta \Delta U} \rangle
\end{equation}
where $\beta = 1/k_{\rm B}T$, $\mu_0^{\rm id}$ is the chemical potential of an ideal gas of particles at the same
temperature and density as the solvent in the mixture, and $\Delta U$ is the change in the potential energy on insertion of a test particle.
The ensemble average is taken over configurations and insertions. We performed $5 \times 10^5$ insertions per
configuration, which we found to be sufficient to obtain a converged value for $\mu_0$.

The chemical potential of the polymers was estimated by the chain increment method\cite{Kumar:1991wa} for both the explicit- and
implicit-solvent models. The incremental chemical potential $\mu^+(M)$ to grow a chain of length $M+1$
from a chain of length $M$ is
\begin{equation}
\beta \mu^+(M) = -\ln \langle e^{-\beta \Delta U} \rangle
\end{equation}
where the ensemble average is now taken over configurations and test insertions onto the ends of chains of length $M$.
The incremental chemical potential $\mu^+(M)$ converges to a constant value $\mu^+(\infty)$ for sufficiently large $M$.\cite{Kumar:1991wa}
The greatest deviations from $\mu^+(\infty)$ occur for a single monomer, $\mu^+(0)$, which can instead be measured by
Widom insertion.\cite{Widom:1963gu} The chemical potential of the chain was then approximated as\cite{Sheng:1995tv}
\begin{equation}
\mu_i = \mu_i^{\rm id} + \mu^+(0) + (M_i-1) \mu^+(\infty)
\end{equation}
where $\mu_i^{\rm id}$ is the chemical potential of an ideal gas of chains of length $M_i$ at the same chain density
and temperature as in the mixture.

In order to improve convergence of the ensemble averages, test particles were inserted onto the chain ends at
positions consistent with the bond length distribution.\cite{Sheng:1995tv,Smit:1992kg} The test particle position was drawn with random orientation
relative to the chain end at distance $r$  distributed according to $r^2 \exp(-\beta \phi(r))$ in the range $0.7\,\sigma \le r \le 1.3\,\sigma$,
where $\phi(r)$ is the interaction potential between bonded monomers
given by eqs.~(\ref{eq:wca}) and (\ref{eq:FENE}). An additional analytical contribution to the
chemical potential was then included to account for the weighted sampling.\cite{Sheng:1995tv} To measure $\mu^+(10)$ and $\mu^+(80)$, we
performed 100 trial insertions per chain end in each configuration and, finding that
$\mu^+(10) \approx \mu^+(80)$ within statistical uncertainty, took $\mu^+(\infty) \approx \mu^+(10)$. The monomer excess chemical potential, $\mu^+(0)$, was measured using $5 \times 10^5$
random insertions per configuration, as for the solvent.

In our implicit-solvent theoretical model (eq.~\ref{eq:migrate}),\cite{Howard:2017bq,Howard:2017vu} the diffusive driving force was given by the gradient
of $\mu_i$. However, this expression must be modified to account for the presence of the solvent in the explicit model.
If the mixture is approximately incompressible, the solvent and polymer concentrations are not independent. Assuming that the
the volume occupied by one polymer chain of length $M_i$ is roughly equal to $M_i$ solvent particles, the chemical
potential $\mu_i$ is replaced in eq.~\ref{eq:migrate} by the \textit{exchange} chemical
potential $\Delta \mu_i = \mu_i - M_i \mu_0$.\cite{Schaefer:2016cc,Batchelor:1976gz} For the implicit model, the solvent is effectively incorporated
into the free energy of the polymers, giving $\Delta \mu_i = \mu_i$ and recovering eq.~\ref{eq:migrate} as expected.
Batchelor showed that the diffusion process resulting from gradients of $\Delta \mu_i$ is equivalent to
sedimentation of the solute (polymers) in a force-free solvent.\cite{Batchelor:1976gz,Batchelor:1983jm}

\begin{figure}[h]
	\centering
	\includegraphics[width=7.5cm]{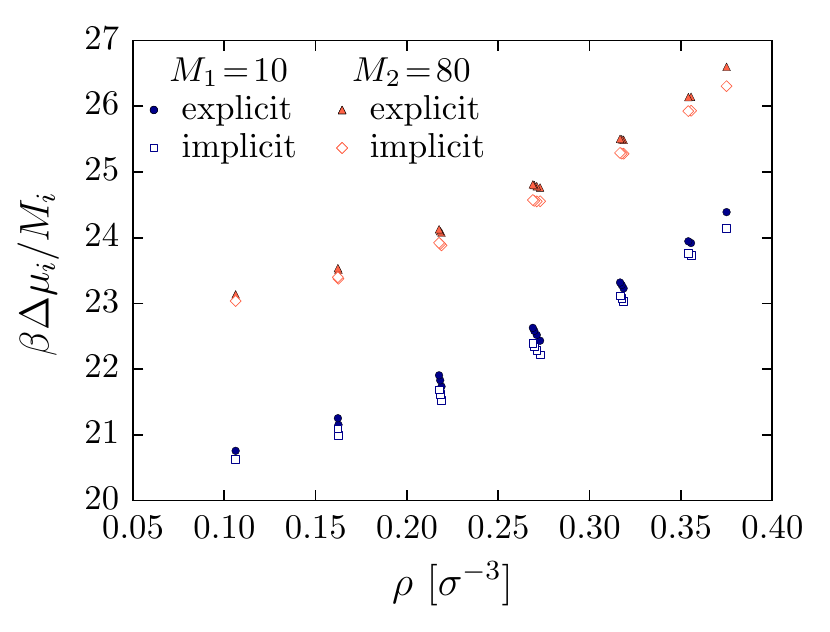}
	\caption{Exchange chemical potential per monomer, $\beta \Delta \mu_i / M_i$, for the short ($M_1 = 10$) and long ($M_2 = 80$)
	         polymers as a function of total monomer density $\rho$ for the explicit- and implicit-solvent models. Each point
	         corresponds to a different mixture composition (see Supplementary Material).}
	\label{fig:chempot}
\end{figure}

Although $\Delta \mu_i$ was determined across a wide range of mixture compositions (Figure~\ref{fig:chempot}),
we found that it was essentially only a function of the total monomer density $\rho$. As expected, $\Delta \mu_i$
increased for larger $\rho$. This is in qualitative agreement with our analysis of stratification in mixtures of
hard-sphere chains, where the excess chemical potential per monomer depended primarily on the total monomer packing
fraction and more weakly on the total chain number density.\cite{Howard:2017vu} Most importantly, $\Delta \mu_i$ is in
excellent agreement between the two models for both the short and long polymers. We then expect
both the explicit- and implicit-solvent models to give equivalent driving forces for diffusion and stratification.

\subsection{Equilibrium diffusion coefficient}
Having found good agreement in the exchange chemical potential between the two models, we considered dynamic effects
in eq.~\ref{eq:migrate}. The polymer mobility $\mathcal{M}_i$ relates the diffusive driving force on the polymers to the
migration velocity. In our theoretical model, we assumed that $\mathcal{M}_i$ could be estimated from the
equilibrium diffusion coefficient $D_i$. In our simulations, $D_i$ for the implicit-solvent model was matched to the
explicit-solvent model in the limit of infinite dilution, but the models may deviate at higher concentrations.
Differences in the concentration-dependence of $D_i$, and hence $\mathcal{M}_i$, may accordingly influence the relative
migration speeds of the components during drying.

We measured $D_i$ for both the explicit- and implicit-solvent models across a wide range of concentrations and compositions
from the mean-squared displacement of the polymer centers of mass. The values of all measured diffusion coefficients are
available in the Supplementary Material. At low total monomer densities, the agreement between the two models was good,
as expected from the model fitting, with larger discrepancies at higher monomer densities. Over the range of
compositions relevant to the drying simulations, the agreement between the two models was quite good, with a deviation of
at most 20\% for the short polymers and 30\% for the long polymers relative to the explicit-solvent model.

\begin{figure}[h]
	\centering
	\includegraphics[width=7.8cm]{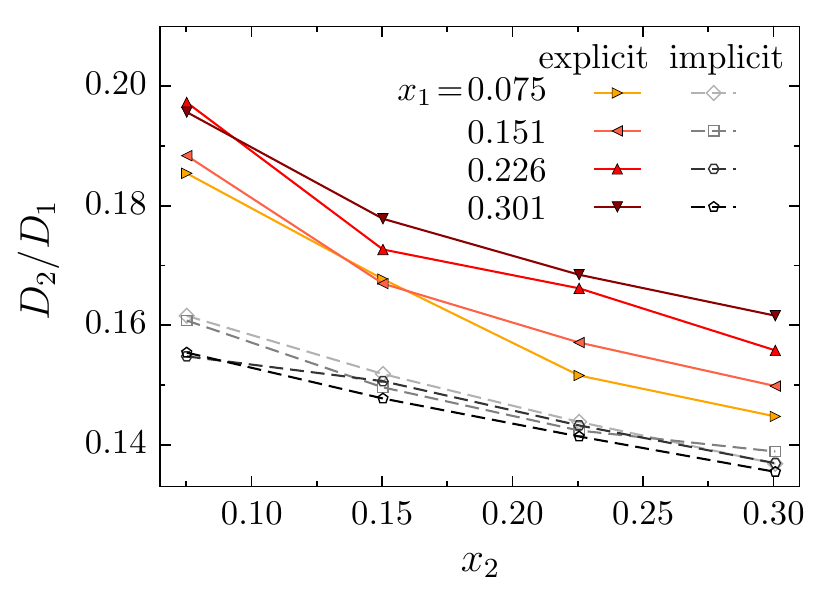}
	\caption{Relative diffusion constants $D_2 / D_1$ as a function of mass fraction $x_2$ of long polymers.
	         Each line represents a different mass fraction for the short polymers $x_1$. For the implicit model,
	         shown with dashed lines, we used the mass fractions the system would have if the solvent particles were included
	         to facilitate comparison with the explicit model.}
	\label{fig:ratio_diffusion}
\end{figure}

According to eq.~\ref{eq:migrate}, a larger value of $\mathcal{M}_2/\mathcal{M}_1$, and so approximately $D_2/D_1$,
should result in a larger difference in migration velocities for the same chemical potential gradient and lead to stronger stratification.
For all concentrations and compositions investigated, $D_2/D_1$ was larger for the
explicit-solvent model (Figure~\ref{fig:ratio_diffusion}). This suggests that stronger stratification
might be expected for the explicit model on the basis of the diffusion coefficients, which is the opposite of the observed behavior. Accordingly, we concluded that
differences in the equilibrium diffusion coefficients were not likely the reason for lack of stratification in the explicit-solvent model.

\subsection{Hydrodynamic interactions~\label{sec:hydrodynamics}}
Given that temperature gradients, chemical potentials, and equilibrium diffusion coefficients do not explain the
difference in stratification in the implicit- and explicit-solvent models, it remains now to test the influence of
hydrodynamic interactions. Hydrodynamic interactions are inherently present in the explicit-solvent model,
but are absent from the implicit-solvent Langevin dynamics simulations. The recent analysis by Sear and
Warren demonstrated how solvent backflow reduces the phoretic motion of a large colloid in an ideal polymer
solution.\cite{Sear:2017cv} Brady showed from a rigorous microscopic perspective how this fundamentally results from a neglect of
hydrodynamic interactions, which modify the mobility.\cite{Brady:2011bh}

\begin{figure}[h]
	\centering
	\includegraphics[width=7.5cm]{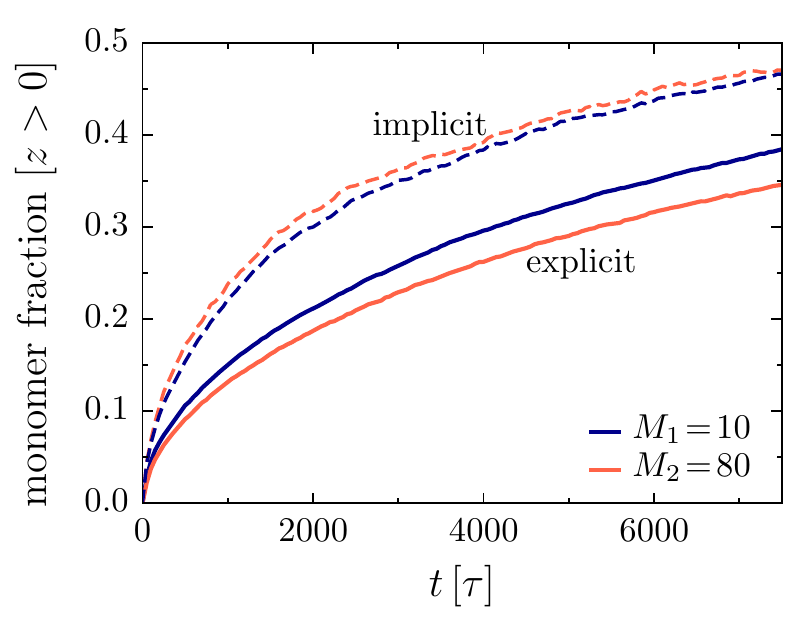}
	\caption{Fraction of monomers for short ($M_1 = 10$) and long ($M_2 = 80$) chains diffusing into the
	        with $z > 0$ over time. Initially, all monomers have $z < 0$. Solid line is
            the explicit solvent model, while dashed lines are the implicit solvent model.}
	\label{fig:polymer_flow}
\end{figure}

We performed a simple simulation to test the effect of hydrodynamic interactions on polymer migration. We constructed a
simulation box with $L = 40\,\sigma$ and $L_{\rm z} = 80\,\sigma$. Half of the box ($z > 0$) was filled with
pure solvent, while the other half ($z < 0$) was filled with a solution of either long or short polymers. A semipermeable membrane
at $z=0\,\sigma$ separated the two compartments, and walls were placed at $z=\pm 40\,\sigma$. Interactions with the walls
were given by eq.~\ref{eq:wall} truncated at distances greater than $3\,\sigma$ with $\varepsilon_{\rm w} = 1\,\varepsilon$, while the
monomer interactions with the membrane were modeled by eq.~\ref{eq:wall} truncated at $(2/5)^{1/6}\,\sigma$ with $\varepsilon_{\rm w} = 1\,\varepsilon$.
The solvent was allowed to exchange across the membrane, coming
to equal chemical potential, but the polymers were confined to $z < 0\,\sigma$. For the explicit-solvent model, there were
78720 solvent particles in the box and the initial monomer density of the polymers in the
compartment was $\rho = 0.2/\sigma^3$ (12800 monomers). The polymers in the explicit solvent
were used as starting configurations for the implicit-solvent model.

We subsequently removed the membrane and allowed the polymers to diffuse.
Based on Figure~\ref{fig:chempot}, the initial gradient in $\Delta \mu_i$ across the membrane should be comparable for
both models, and so any differences in the diffusive flux should be due to hydrodynamic interactions through the mobility.
Figure~\ref{fig:polymer_flow} shows the fraction of monomers which
moved into the opposite side of the box ($z > 0\,\sigma$) at a given time, averaged over 25 independent simulations.
At steady state, half of the monomers should have $z > 0\,\sigma$. It is clear that the dynamics of the
implicit-solvent model are faster than the explicit-solvent model, in agreement with the considerations in
refs.~\citenum{Sear:2017cv} and~\citenum{Brady:2011bh}. Moreover, the long ($M_2 = 80$) polymers and short ($M_1 = 10$) polymers
responded differently to the same density gradient between the two models. In the implicit-solvent model, the long
polymers migrated \textit{faster} than the short polymers, consistent with eq.~\ref{eq:migrate} and stratification in
the drying simulations. On the other hand, the long polymers migrated much \textit{slower} than the short polymers in
the explicit-solvent model, explaining the lack of stratification in the drying simulations.

\section{Conclusions \label{sec:conclude}}
We demonstrated the influence of hydrodynamic interactions on stratification in a drying polymer mixture using
explicit-solvent and implicit-solvent computer simulations. The implicit-solvent model predicted stratification
in agreement with previous simulations and theoretical considerations.\cite{Howard:2017vu} However, no such stratification was found for
the explicit-solvent model at the drying conditions considered. Despite good mapping of the equilibrium bulk properties
(chemical potential, diffusion coefficient) between the explicit- and implicit-solvent models, hydrodynamic
interactions out of equilibrium were shown to alter the polymer diffusion in a way that is consistent with the
lack of stratification. Our analysis directly tests and confirms that implicit-solvent simulations and
theoretical models lacking hydrodynamic interactions are not capable of quantitatively predicting stratification,
in agreement with the analysis of Sear and Warren\cite{Sear:2017cv} for the special case of a large
colloid in an ideal polymer solution. In future, hydrodynamic interactions must be incorporated into any simulations aiming to study
stratification, e.g., through explicit-solvent molecular dynamics or with an appropriate mesoscale simulation method.

While there are currently no experiments available for stratification in polymer mixtures, colloid mixtures have been shown to stratify in
experiments.\cite{Fortini:2016ip,MartinFabiani:2016fj,Makepeace:2017ht} The extent of colloid stratification in the experiments
appears weaker than predicted by models lacking hydrodynamic interactions,\cite{Fortini:2016ip} consistent with our analysis.
Larger chemical potential gradients on the larger component would be required to increase the stratification, which could
be induced by, for example, larger size ratios or additional cross-interactions between the components.

\section*{Supplementary Material}
See supplementary material for single-chain structure for both models and equilibrium bulk properties of mixtures.

\begin{acknowledgments}
We gratefully acknowledge use of computational resources supported by the Princeton Institute for Computational Science
and Engineering (PICSciE) and the Office of Information Technology's High Performance Computing Center and Visualization
Laboratory at Princeton University. Financial support for this work was provided by the Princeton Center for Complex
Materials, a U.S. National Science Foundation Materials Research Science and Engineering Center (award DMR-1420541).
\end{acknowledgments}

\bibliography{stratification_explicit}

\end{document}